\pdfoutput=1

\documentclass[12pt,a4paper]{article}

\usepackage{ifthen} % for conditional statements
\newboolean{pdflatex}
\setboolean{pdflatex}{true} % False for eps figures 

\newboolean{articletitles}
\setboolean{articletitles}{true} % False removes titles in references

\newboolean{uprightparticles}
\setboolean{uprightparticles}{false} %True for upright particle symbols

\newboolean{inbibliography}
\setboolean{inbibliography}{false} %True once you enter the bibliography

\usepackage{microtype}
\usepackage{overpic}
\usepackage{rotating}
\usepackage{lineno}  % for line numbering during review
\usepackage{xspace} % To avoid problems with missing or double spaces after
\usepackage{caption} %these three command get the figure and table captions automatically small

\usepackage{graphicx}  % to include figures (can also use other packages)
\DeclareGraphicsExtensions{.pdf,.png,.jpg}
\usepackage{color}
\usepackage{colortbl}
\graphicspath{{./figs/}} % Make Latex search fig subdir for figures

\usepackage{amsmath} % Adds a large collection of math symbols
\usepackage{amssymb}
\usepackage{amsfonts}
\usepackage{upgreek} % Adds in support for greek letters in roman typeset
\usepackage{acronym}

\newcommand*\patchAmsMathEnvironmentForLineno[1]{%
\expandafter\let\csname old#1\expandafter\endcsname\csname #1\endcsname
\expandafter\let\csname oldend#1\expandafter\endcsname\csname
end#1\endcsname
 \renewenvironment{#1}%
   {\linenomath\csname old#1\endcsname}%
   {\csname oldend#1\endcsname\endlinenomath}%
}
\newcommand*\patchBothAmsMathEnvironmentsForLineno[1]{%
  \patchAmsMathEnvironmentForLineno{#1}%
  \patchAmsMathEnvironmentForLineno{#1*}%
}
\AtBeginDocument{%
\patchBothAmsMathEnvironmentsForLineno{equation}%
\patchBothAmsMathEnvironmentsForLineno{align}%
\patchBothAmsMathEnvironmentsForLineno{flalign}%
\patchBothAmsMathEnvironmentsForLineno{alignat}%
\patchBothAmsMathEnvironmentsForLineno{gather}%
\patchBothAmsMathEnvironmentsForLineno{multline}%
\patchBothAmsMathEnvironmentsForLineno{eqnarray}%
}

\usepackage{hyperref}    % Hyperlinks in references
\usepackage[all]{hypcap} % Internal hyperlinks to floats.

\def\lhcb {\mbox{LHCb}\xspace}

\def\babar  {\mbox{BaBar}\xspace}
\def\belle  {\mbox{Belle}\xspace}

\def\MagUp {\mbox{\em Mag\kern -0.05em Up}\xspace}

\ifthenelse{\boolean{uprightparticles}}%
{

 \def\Pmu         {\ensuremath{\upmu}\xspace}                 
 \def\Pnu         {\ensuremath{\upnu}\xspace}                 
                  
 \def\Ppi         {\ensuremath{\uppi}\xspace}

 \def\Ptau        {\ensuremath{\uptau}\xspace}

 \def\Ppsi        {\ensuremath{\uppsi}\xspace}

 \def\PDelta      {\ensuremath{\Delta}\xspace}                 
 \def\PXi      {\ensuremath{\Xi}\xspace}                 
 \def\PLambda      {\ensuremath{\Lambda}\xspace}                 
 \def\PSigma      {\ensuremath{\Sigma}\xspace}                 
 \def\POmega      {\ensuremath{\Omega}\xspace}                 
 \def\PUpsilon      {\ensuremath{\Upsilon}\xspace}                 
 
 \def\PB      {\ensuremath{\mathrm{B}}\xspace}                 
                  
 \def\PD      {\ensuremath{\mathrm{D}}\xspace}

 \def\PJ      {\ensuremath{\mathrm{J}}\xspace}                 
 \def\PK      {\ensuremath{\mathrm{K}}\xspace}

 \def\Pb      {\ensuremath{\mathrm{b}}\xspace}                 
 \def\Pc      {\ensuremath{\mathrm{c}}\xspace}

 \def\Pi      {\ensuremath{\mathrm{i}}\xspace}

 \def\Ps      {\ensuremath{\mathrm{s}}\xspace}

}
{

 \def\Pmu         {\ensuremath{\mu}\xspace}                 
 \def\Pnu         {\ensuremath{\nu}\xspace}                 
                  
 \def\Ppi         {\ensuremath{\pi}\xspace}

 \def\Ptau        {\ensuremath{\tau}\xspace}

 \def\Ppsi        {\ensuremath{\psi}\xspace}                 
                  
 \mathchardef\PDelta="7101
 \mathchardef\PXi="7104
 \mathchardef\PLambda="7103
 \mathchardef\PSigma="7106
 \mathchardef\POmega="710A
 \mathchardef\PUpsilon="7107
                  
 \def\PB      {\ensuremath{B}\xspace}                 
                  
 \def\PD      {\ensuremath{D}\xspace}

 \def\PJ      {\ensuremath{J}\xspace}                 
 \def\PK      {\ensuremath{K}\xspace}

 \def\Pb      {\ensuremath{b}\xspace}                 
 \def\Pc      {\ensuremath{c}\xspace}

 \def\Pi      {\ensuremath{i}\xspace}

 \def\Ps      {\ensuremath{s}\xspace}

}

\makeatletter
\ifcase \@ptsize \relax% 10pt
  \newcommand{\miniscule}{\@setfontsize\miniscule{4}{5}}% \tiny: 5/6
\or% 11pt
  \newcommand{\miniscule}{\@setfontsize\miniscule{5}{6}}% \tiny: 6/7
\or% 12pt
  \newcommand{\miniscule}{\@setfontsize\miniscule{5}{6}}% \tiny: 6/7
\fi
\makeatother

\DeclareRobustCommand{\optbar}[1]{\shortstack{{\miniscule (\rule[.5ex]{1.25em}{.18mm})}
  \\ [-.7ex] $#1$}}

\def\mup        {{\ensuremath{\Pmu^+}}\xspace}
\def\mun        {{\ensuremath{\Pmu^-}}\xspace} % muon negative (\mum is taken)

\def\taum       {{\ensuremath{\Ptau^-}}\xspace}

\def\neu        {{\ensuremath{\Pnu}}\xspace}
\def\neub       {{\ensuremath{\overline{\Pnu}}}\xspace}
\def\neum       {{\ensuremath{\neu_\mu}}\xspace}
\def\neumb      {{\ensuremath{\neub_\mu}}\xspace}
\def\neut       {{\ensuremath{\neu_\tau}}\xspace}
\def\neutb      {{\ensuremath{\neub_\tau}}\xspace}
\def\squark    {{\ensuremath{\Ps}}\xspace}

\def\cquark    {{\ensuremath{\Pc}}\xspace}

\def\bquark    {{\ensuremath{\Pb}}\xspace}

\def\pion   {{\ensuremath{\Ppi}}\xspace}

\def\pip    {{\ensuremath{\pion^+}}\xspace}
\def\pim    {{\ensuremath{\pion^-}}\xspace}

\def\kaon    {{\ensuremath{\PK}}\xspace}
  \def\Kbar    {{\kern 0.2em\overline{\kern -0.2em \PK}{}}\xspace}

\def\KorKbar    {\kern 0.18em\optbar{\kern -0.18em K}{}\xspace}

\def\Kp      {{\ensuremath{\kaon^+}}\xspace}
\def\Km      {{\ensuremath{\kaon^-}}\xspace}

\def\KS      {{\ensuremath{\kaon^0_{\rm\scriptscriptstyle S}}}\xspace}

  \def\Dbar    {{\kern 0.2em\overline{\kern -0.2em \PD}{}}\xspace}
\def\D       {{\ensuremath{\PD}}\xspace}

\def\DorDbar    {\kern 0.18em\optbar{\kern -0.18em D}{}\xspace}
\def\Dz      {{\ensuremath{\D^0}}\xspace}

\def\Dstar   {{\ensuremath{\D^*}}\xspace}

\def\Dstarp  {{\ensuremath{\D^{*+}}}\xspace}

\def\B       {{\ensuremath{\PB}}\xspace}
\def\Bbar    {{\ensuremath{\kern 0.18em\overline{\kern -0.18em \PB}{}}}\xspace}
\def\Bb      {{\ensuremath{\Bbar}}\xspace}
\def\BorBbar    {\kern 0.18em\optbar{\kern -0.18em B}{}\xspace}

\def\Bzb     {{\ensuremath{\Bbar{}^0}}\xspace}
\def\Bu      {{\ensuremath{\B^+}}\xspace}

\def\Bp      {{\ensuremath{\Bu}}\xspace}

\def\Bsb     {{\ensuremath{\Bbar{}^0_\squark}}\xspace}

\def\jpsi     {{\ensuremath{{\PJ\mskip -3mu/\mskip -2mu\Ppsi\mskip 2mu}}}\xspace}

  \def\Y#1S{\ensuremath{\PUpsilon{(#1S)}}\xspace}% no space before {...}!

\def\Lz          {{\ensuremath{\PLambda}}\xspace}
\def\Lbar        {{\ensuremath{\kern 0.1em\overline{\kern -0.1em\PLambda}}}\xspace}
\def\LorLbar    {\kern 0.18em\optbar{\kern -0.18em \PLambda}{}\xspace}

\def\to                 {\ensuremath{\rightarrow}\xspace}

\def\AT#1     {\ensuremath{A_{\mathrm{T}}^{#1}}\xspace}           % 2

\def\C#1      {\ensuremath{\mathcal{C}_{#1}}\xspace}                       % 9
\def\Cp#1     {\ensuremath{\mathcal{C}_{#1}^{'}}\xspace}                    % 7
\def\Ceff#1   {\ensuremath{\mathcal{C}_{#1}^{\mathrm{(eff)}}}\xspace}        % 9  
\def\Cpeff#1  {\ensuremath{\mathcal{C}_{#1}^{'\mathrm{(eff)}}}\xspace}       % 7
\def\Ope#1    {\ensuremath{\mathcal{O}_{#1}}\xspace}                       % 2
\def\Opep#1   {\ensuremath{\mathcal{O}_{#1}^{'}}\xspace}                    % 7

\newcommand{\tev}{\ifthenelse{\boolean{inbibliography}}{\ensuremath{~T\kern -0.05em eV}\xspace}{\ensuremath{\mathrm{\,Te\kern -0.1em V}}}\xspace}
\newcommand{\gev}{\ensuremath{\mathrm{\,Ge\kern -0.1em V}}\xspace}
\newcommand{\mev}{\ensuremath{\mathrm{\,Me\kern -0.1em V}}\xspace}
\newcommand{\kev}{\ensuremath{\mathrm{\,ke\kern -0.1em V}}\xspace}
\newcommand{\ev}{\ensuremath{\mathrm{\,e\kern -0.1em V}}\xspace}
\newcommand{\gevc}{\ensuremath{{\mathrm{\,Ge\kern -0.1em V\!/}c}}\xspace}
\newcommand{\mevc}{\ensuremath{{\mathrm{\,Me\kern -0.1em V\!/}c}}\xspace}
\newcommand{\gevcc}{\ensuremath{{\mathrm{\,Ge\kern -0.1em V\!/}c^2}}\xspace}
\newcommand{\gevgevcccc}{\ensuremath{{\mathrm{\,Ge\kern -0.1em V^2\!/}c^4}}\xspace}
\newcommand{\mevcc}{\ensuremath{{\mathrm{\,Me\kern -0.1em V\!/}c^2}}\xspace}

\def\mum  {\ensuremath{{\,\upmu\rm m}}\xspace}

\def\invfb   {\ensuremath{\mbox{\,fb}^{-1}}\xspace}

\newcommand{\stat}{\ensuremath{\mathrm{\,(stat)}}\xspace}
\newcommand{\syst}{\ensuremath{\mathrm{\,(syst)}}\xspace}

\def\gsim{{~\raise.15em\hbox{$>$}\kern-.85em
          \lower.35em\hbox{$\sim$}~}\xspace}
\def\lsim{{~\raise.15em\hbox{$<$}\kern-.85em
          \lower.35em\hbox{$\sim$}~}\xspace}

\def\ptot       {\mbox{$p$}\xspace}
\def\pt         {\mbox{$p_{\rm T}$}\xspace}

\def\evtgen     {\mbox{\textsc{EvtGen}}\xspace}

\def\geant      {\mbox{\textsc{Geant4}}\xspace}

\def\photos     {\mbox{\textsc{Photos}}\xspace}

\def\pythia     {\mbox{\textsc{Pythia}}\xspace}

\def\tell1  {TELL1\xspace}
\def\ukl1   {UKL1\xspace}

\newcommand{\ie}{\mbox{\itshape i.e.}\xspace}

\def\RDst {\ensuremath{\mathcal{R}(\Dstar)}\xspace}

\def\mmsq {\ensuremath{m^2_{\rm miss}}\xspace}
\def\El {\ensuremath{E^{*}_{\mu}}\xspace}
\def\qq {\ensuremath{q^{2}}\xspace}

\def\sigornorm {\ensuremath{\Dstarp \mun\xspace}}
\def\normmode {\ensuremath{\Bzb\to~\Dstarp\mun\neumb}\xspace}
\def\sigmode {\ensuremath{\Bzb\to~\Dstarp\taum\neutb }\xspace}
\def\sigtomu {\ensuremath{\Bzb\to~\Dstarp\taum(\to~\mun\neut\neumb)\neutb }\xspace}

\usepackage{cite} % Allows for ranges in citations
\usepackage{mciteplus}

\usepackage{longtable} % only for template; not usually to be used in PAPERs

\begin{document}

%%%%%%%%%%%%%%%%%%%%%%%%%
%%%%% Title     %%%%%%%%%
%%%%%%%%%%%%%%%%%%%%%%%%%
\renewcommand{\thefootnote}{\fnsymbol{footnote}}
\setcounter{footnote}{1}

% $Id: title-LHCb-PAPER.tex 56951 2014-06-30 13:45:01Z roldeman $
% ===============================================================================
% Purpose: LHCb-PAPER journal paper title page template
% Author: 
% Created on: 2010-09-25
% ===============================================================================

%%%%%%%%%%%%%%%%%%%%%%%%%
%%%%%  TITLE PAGE  %%%%%%
%%%%%%%%%%%%%%%%%%%%%%%%%
\begin{titlepage}
\pagenumbering{roman}

% Header ---------------------------------------------------
\vspace*{-1.5cm}
\centerline{\large EUROPEAN ORGANIZATION FOR NUCLEAR RESEARCH (CERN)}
\vspace*{1.5cm}
\hspace*{-0.5cm}
\begin{tabular*}{\linewidth}{lc@{\extracolsep{\fill}}r}
\ifthenelse{\boolean{pdflatex}}% Logo format choice
{\vspace*{-2.7cm}\mbox{\!\!\!\includegraphics[width=.14\textwidth]{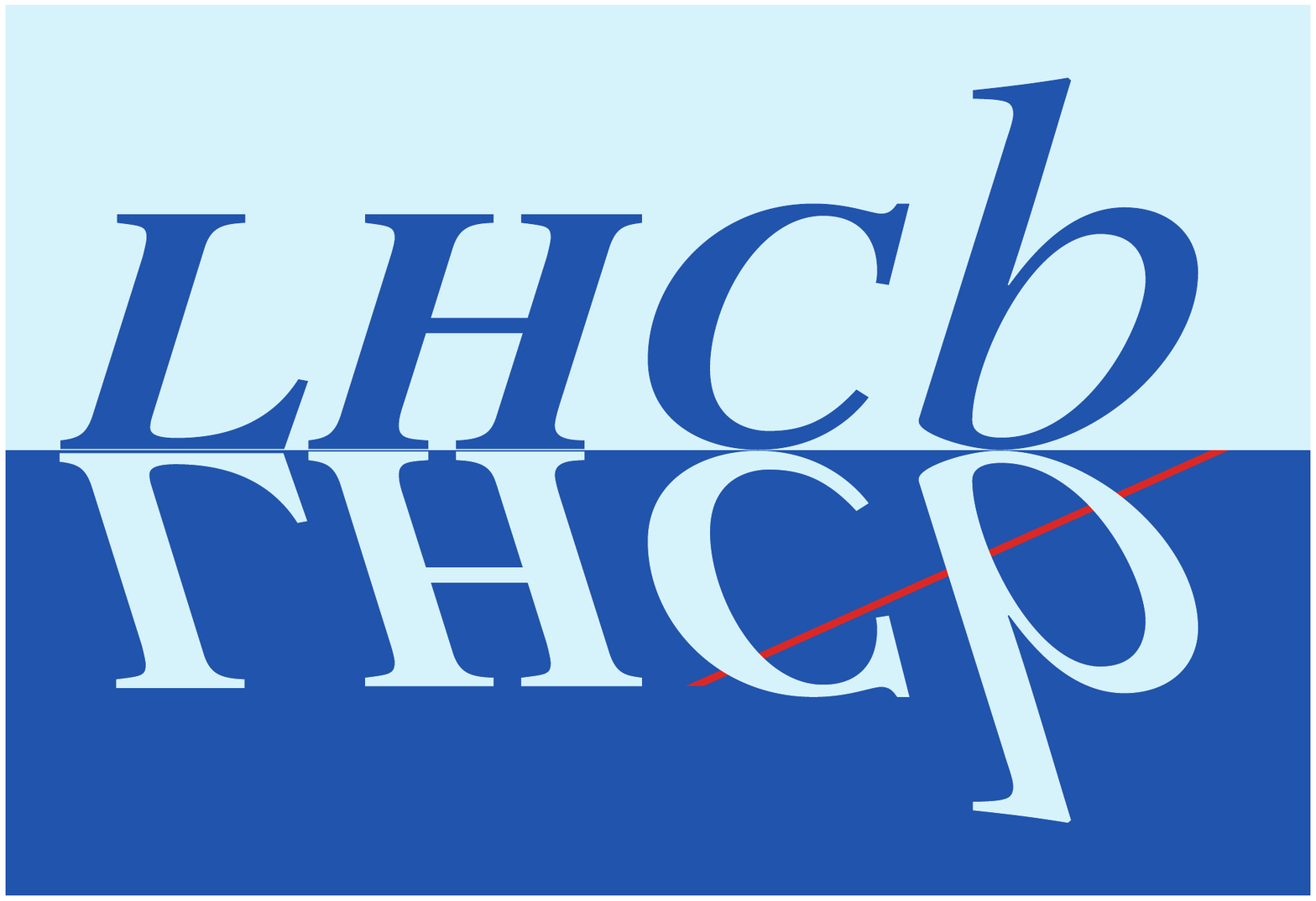}} & &}%
{\vspace*{-1.2cm}\mbox{\!\!\!\includegraphics[width=.12\textwidth]{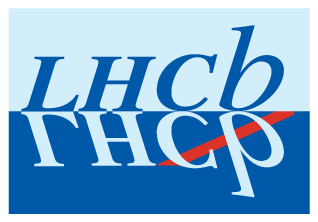}} & &}%
\\
 & & CERN-PH-EP-2015-150 \\  % ID 
 & & LHCb-PAPER-2015-025 \\  % ID 
 & & June 29, 2015 \\%\today \\ % Date - Can also hardwire e.g.: 23 March 2010
 & & \\
% not in paper \hline
\end{tabular*}

\vspace*{3.2cm}

% Title --------------------------------------------------
{\bf\boldmath\LARGE
\begin{center}
Measurement of the ratio of \\ branching fractions $\mathcal{B}(\Bzb \to \Dstarp\taum\neutb)/\mathcal{B}(\Bzb \to \Dstarp\mun\neumb)$
\end{center}
}

\vspace*{2.0cm}
%\vspace*{1.85cm}

% Authors -------------------------------------------------
\begin{center}
%In the footnote, replace 'paper' by 'letter' in case of submission to PRL or PLB 
The LHCb collaboration\footnote{Authors are listed at the end of this letter.}
\end{center}

\vspace{\fill}

% Abstract -----------------------------------------------
\begin{abstract}
  \noindent
The branching fraction ratio 
$\RDst\!\equiv\!\mathcal{B}(\Bzb\!\!\to\!\!D^{*+}\taum\neutb)/\mathcal{B}(\Bzb\!\!\to\!\!D^{*+}\mun\neumb)$ 
is measured using a sample of proton-proton collision data corresponding to 3.0\invfb of integrated luminosity recorded by the LHCb experiment during 2011 and 2012.  The tau lepton is identified in the decay mode 
 $\taum \to \mun\neumb\neut$. The semitauonic decay is sensitive 
to contributions from non-Standard-Model particles that preferentially couple to the third generation of fermions, in 
particular \mbox{Higgs-like} charged scalars. A multidimensional fit to kinematic distributions of the candidate $\Bzb$ decays gives $\RDst = 0.336\, \pm 0.027\stat\, \pm 0.030\syst$.
This result, which is the first measurement of this quantity at a hadron collider, is $2.1$ standard deviations larger than the value expected from lepton universality in the Standard Model.

\end{abstract}

%\vspace*{2.0cm}
\vspace{1.8cm}
\begin{center}
  Published in Phys.~Rev.~Lett. {\bf 115}, 111803 (2015)
\end{center}

\vspace{\fill}

{\footnotesize 
\centerline{\copyright~CERN on behalf of the \lhcb collaboration, license \href{http://creativecommons.org/licenses/by/4.0/}{CC-BY-4.0}.}}
\vspace*{2mm}

\end{titlepage}

%%%%%%%%%%%%%%%%%%%%%%%%%%%%%%%%
%%%%%  EOD OF TITLE PAGE  %%%%%%
%%%%%%%%%%%%%%%%%%%%%%%%%%%%%%%%

%  empty page follows the title page ----
\newpage
\setcounter{page}{2}
\mbox{~}

\cleardoublepage

\renewcommand{\thefootnote}{\arabic{footnote}}
\setcounter{footnote}{0}

%%%%%%%%%%%%%%%%%%%%%%%%%

\pagestyle{plain} % restore page numbers for the main text
\setcounter{page}{1}
\pagenumbering{arabic}

%\linenumbers

Lepton universality, enshrined within the Standard Model (SM), requires equality of couplings between the gauge bosons
and the three families of leptons. Hints of lepton non-universal effects in $\Bp \to \Kp e^{+}e^{-}$ 
and $\Bp \to \Kp \mup\mun$ decays \cite{LHCb-PAPER-2014-024} have been seen, but no definitive observation of a deviation has 
yet been made. However, a large class of models that extend the SM contain additional 
interactions involving enhanced couplings to the third generation that would violate this principle. 
Semileptonic decays of $b$ hadrons (particles containing a $b$ quark) to third generation leptons provide
a sensitive probe for such effects. 
In particular, the presence of additional charged Higgs bosons, which are often required in these models, can have a significant effect on the rate of the semitauonic decay \sigmode \cite{Tanaka:1994ay}. The use of charge-conjugate modes is implied throughout this Letter.

Semitauonic \B meson decays have been observed by the \babar and Belle collaborations \cite{Matyja:2007kt,Aubert:2007dsa,Bozek:2010xy,Lees:2012xj,Lees:2013uzd}. 
Recently \babar reported updated measurements \cite{Lees:2012xj,Lees:2013uzd} of the ratios of branching fractions, 
$\RDst\equiv \mathcal{B}(\Bzb\to~D^{*+}\taum\neutb)/\mathcal{B}(\Bzb\to~D^{*+}\mun\neumb)$ and  
$\mathcal{R}(D)\equiv \mathcal{B}(\Bzb\to~D^{+}\taum\neutb)/\mathcal{B}(\Bzb\to~D^{+}\mun\neumb)$, which show 
deviations of $2.7\sigma$ and $2.0\sigma$, respectively, from the SM predictions \cite{Fajfer:2012dp,Bailey:2012jg}.
These ratios
have been calculated to high precision, owing to the cancellation of most of the uncertainties associated with the strong interaction in the $\B$ to $D^{(*)}$ transition. Within the SM they differ from unity mainly because of 
phase-space effects due to the differing charged lepton masses. 

This Letter presents the first measurement of $\RDst$ in hadron collisions
using
the data recorded by
the LHCb detector at the Large Hadron Collider in 2011--2012. The data correspond to integrated 
luminosities of 1.0\invfb and 2.0\invfb, collected at proton-proton ($pp$) center-of-mass energies of 7 TeV and 8 TeV, respectively. 
The \sigmode decay with $\taum\to\mun\neumb\neut$ (the signal channel) and the \normmode decay (the normalization channel) produce identical visible final-state topologies; consequently both are selected by a common reconstruction procedure. 
The selection identifies semileptonic \Bzb decay candidates containing a muon candidate and a $\Dstarp$ candidate reconstructed through the decay
chain $D^{*+}\rightarrow D^0(\to \Km \pip)\pi^+$.
The selected sample contains contributions from the signal and  
the normalization channel, as well as several background processes, which include partially reconstructed 
\B decays and candidates from combinations of unrelated particles from different $b$ hadron decays.  
The kinematic and topological properties of the various components are exploited to
suppress the background contributions.  Finally, the signal, the normalization component and the residual background are statistically disentangled  with 
a multidimensional fit to the data using template distributions derived from control samples or from simulation validated against data.

\label{sec:Detector}

The \lhcb detector~\cite{Alves:2008zz,LHCb-DP-2014-002} is a single-arm forward
spectrometer covering the \mbox{pseudorapidity} range $2<\eta <5$,
designed for the study of particles containing \bquark or \cquark
quarks. The detector includes a high-precision tracking system
consisting of a silicon-strip vertex detector surrounding the $pp$
interaction region~\cite{LHCb-DP-2014-001}, a large-area silicon-strip detector located
upstream of a dipole magnet with a bending power of about
$4{\rm\,Tm}$, and three stations of silicon-strip detectors and straw
drift tubes~\cite{LHCb-DP-2013-003} placed downstream of the magnet.
The tracking system provides a measurement of momentum, \ptot, 
of charged particles with
a relative uncertainty that varies from 0.5\% at low momentum to 1.0\% at 200\gevc.
The minimum distance of a track to a primary vertex (PV), the impact parameter, is measured with a resolution of $(15+29/\pt)\mum$,
where \pt is the component of the momentum transverse to the beam, in\,\gevc.
Different types of charged hadrons are distinguished using information
from two ring-imaging Cherenkov detectors~\cite{LHCb-DP-2012-003}. 
Photons, electrons and hadrons are identified by a calorimeter system consisting of
scintillating-pad and preshower detectors, an electromagnetic
calorimeter and a hadronic calorimeter. Muons are identified by a
system composed of alternating layers of iron and multiwire
proportional chambers~\cite{LHCb-DP-2012-002}.
The online event selection is performed by a trigger~\cite{LHCb-DP-2012-004}, 
which consists of a hardware stage, based on information from the calorimeter and muon
systems, followed by a software stage, which applies a full event
reconstruction.

A simulation of $pp$ collisions is provided by
\pythia~\cite{Sjostrand:2006za,Sjostrand:2007gs} 
%(In case only \pythia 6 is used, remove \verb=*Sjostrand:2007gs= from this citation )
 with a specific \lhcb
configuration~\cite{LHCb-PROC-2010-056}.  Decays of hadronic particles
are described by \evtgen~\cite{Lange:2001uf}, in which final-state
radiation is generated using \photos~\cite{Golonka:2005pn}. The
interaction of the generated particles with the detector, and its
response, are implemented using the \geant
toolkit~\cite{Allison:2006ve, Agostinelli:2002hh} as described in
Ref.~\cite{LHCb-PROC-2011-006}.

The trigger requirements are chosen to avoid the imposition of any $\pt$ selection on the muon, or invariant mass requirements on the $\Dstarp \mun$ system, crucial for 
preserving the distinct kinematic distributions of the \sigtomu decay.
Events are required to pass the hardware trigger either because the decay products of the $\Dstarp$ candidate satisfy the hadron trigger requirements or because  high-\pt particles in the event, independent of the $\Dstarp\mun$, satisfy one of the hardware trigger requirements. 
In the software trigger, the events
are required to meet criteria designed to accept $\Dz\to\Km\pip$ candidates 
with $\pt> 2\gevc$.  Quality requirements are applied to the tracks of the charged particles that originate from a candidate \Dz decay: their momenta must exceed 5\gevc and at least 
one must have 
 $\pt> 1.5 \gevc$.  The momentum vector of the \Dz candidate must point back to one of the PVs in 
 the event and the reconstructed mass must be consistent with the known \Dz mass \cite{PDG2014}. 
 
In the offline reconstruction, the $D^0$ candidates satisfying the trigger are further required to have well-identified $\Km$ and $\pip$ daughters, and the decay vertex is required to be significantly separated from any PV. The invariant mass of the \Dz candidate is required to be within 23.5\mevcc of the peak value, corresponding to approximately three times the \Dz mass resolution. These candidates are combined with low-energy pions to form candidate $\Dstarp\to\Dz\pip$ decays, which are 
subjected to a kinematic and vertex fit to the decay chain. Candidates are then required to have a mass difference $\Delta m \equiv m(\Dz\pip)-m(\Dz)$ within 2\mevcc of the known value, corresponding to approximately 2.5 times the observed resolution.   
The muon candidate is required to be consistent with a muon signature in the detector, to have momentum  $3 <p< 100 \gevc$,  
to be significantly separated from the primary vertex, and to form a good vertex with the $D^0$ candidate. The $\Dstarp\mun$ combinations are required to have an invariant
mass less than 5280\mevcc and their momentum vector must point approximately to one of the reconstructed
PV locations, which removes combinatoric candidates while preserving a large fraction of semileptonic decays. In addition to the signal candidates, two independent samples of ``wrong sign" candidates, $\Dstarp\mup$ and $\Dz\pim\mun$, are 
formed for estimating the combinatorial background. The former represents random combinations of $\Dstarp$ candidates with muons from unrelated decays, and the latter is used to model the contribution of misreconstructed $\Dstarp$ decays. Mass regions $5280 < m(\Dstarp\mun) < 10000\mevcc$ and $139 < \Delta m < 160\mevcc$ are included in all samples for study of the combinatorial backgrounds. 
Finally, a sample of candidates is selected where the track paired with the $\Dstarp$ fails all muon identification requirements. These $\Dstarp h^{\pm}$ candidates are used to model the background from hadrons misidentified as muons.

To suppress the contributions of partially reconstructed \B decays, including \B decays to pairs of charmed hadrons, and semileptonic $\Bb \to \Dstarp (\textrm{n}\pi)\mun\neum$ decays with $\textrm{n} \geq 1$ additional pions, the \sigornorm candidates are required to be isolated from additional tracks in the event. 
An algorithm is developed and trained to determine whether a given track is likely to have originated from the signal \B candidate or 
from the rest of the event based on a multivariate analysis (MVA) method. For each track in the event, the algorithm employs information on the track separation from the PV, the track separation from the decay vertex, the 
angle between the track and the candidate momentum vector, the decay length significance of the decay vertex under the hypothesis that the track does not originate from the candidate and the change in this significance under the hypothesis that it does. A signal sample, enriched in \sigmode and \normmode decays, is constructed by requiring that no tracks in the event reach a threshold in the MVA output. In addition, the output is used to select three control samples enriched in partially reconstructed \B decays of interest for  background studies by requiring that only one or two tracks be selected by the MVA ($\Dstarp\mun\pim$ or $\Dstarp\mun\pip\pim$) or that at least one track selected by the MVA passes $K^{\pm}$ identification requirements ($\Dstarp\mun K^{\pm}$). These samples are depleted of \normmode and \sigmode decays and are used to study and constrain the shapes of remaining backgrounds in the signal sample.   

The efficiencies $\varepsilon_{\rm s}$ and $\varepsilon_{\rm n}$ for the signal and the normalization channels, respectively, are determined in simulation. These include the effects of the trigger, event reconstruction, event selection, particle identification procedure, isolation method, and the detector acceptance.
To account for the effect of 
differing detector occupancy distributions between simulation and data, the simulated samples are reweighted to match the occupancy observed in data. The overall efficiency ratio is 
${\varepsilon_{\rm s}}/{\varepsilon_{\rm n}}=(77.6\pm1.4)\%$, with the deviation from unity primarily due to the particle identification, which dominantly removes low-\pt muon candidates, and vertex quality requirements.  

The separation of the signal from the normalization channel, as well as from background processes, is achieved by 
exploiting the distinct kinematic distributions that characterize the various decay modes, resulting from 
the $\mu-\tau$ mass difference and the presence of extra neutrinos from the decay $\taum \to \mun\neumb\neut$.
The most discriminating kinematic variables are the following quantities, computed in the \B rest frame: the muon energy, $E_{\mu}^*$; the missing mass squared, defined as
 $m^2_{\textrm{miss}}=(p^{\mu}_{\B} \!-\! p^{\mu}_{\D} \!-\! p^{\mu}_{\mu})^2$; and the squared four-momentum transfer to the lepton system, $q^2=(p^{\mu}_{\B} - p^{\mu}_{\D})^2$, where $p^{\mu}_{\B}$, $p^{\mu}_{\D}$ and $p^{\mu}_{\mu}$ are the four-momenta of 
 the \B meson, the $D^{*+}$ meson and the muon.  The determination of the rest-frame variables requires knowledge of the \B candidate momentum vector in 
 the laboratory frame, which is estimated from the measured parameters of the reconstructed final-state particles. The \B momentum direction is determined from
 the unit vector to the \B decay vertex from the associated PV.  
 The component of the $B$ momentum along the beam axis is approximated using the relation $(p_{\B})_z=\frac{m_{\B}}{m_{\rm reco}}(p_{\rm reco})_z$, where $m_{\B}$ is the known \B mass, and $m_{\rm reco}$ and $p_{\rm reco}$ are the mass and momentum of the system of reconstructed particles. 
 The rest-frame variables described above are then calculated using the resulting estimated $\B$ four-momentum and the measured four-momenta of the $\mun$ and $\Dstarp$. The rest-frame variables  
are shown in simulation studies to have sufficient resolution ($\approx 15\%$--$20\%$ full width at half maximum) to preserve the discriminating features of the original distributions.

Simulated events are used to derive kinematic distributions from signal and \B backgrounds that are used to fit the data. 
The hadronic transition-matrix elements for \sigmode and \normmode decays are described using form factors derived from heavy quark effective theory~\cite{Caprini:341849}.
Recent world averages
for the corresponding parameters are taken from Ref.~\cite{HFAG}. These values, along with
their correlations and uncertainties, are included as external constraints on the respective fit parameters. The hadronic matrix elements describing \sigmode decays include a helicity-supressed component, which 
is negligible in \normmode decays~\cite{Korner:1990kh}. This parameter is not well-constrained by data; hence, the central value and uncertainty from the sum rule presented
in Ref.~\cite{Fajfer:2012dp} are used as a constraint. It is assumed that the kinematic properties of the \sigmode decay are not modified by any SM extensions. 

For the background semileptonic decays $\Bb \to (D^{}_{1}(2420),D^{*}_{2}(2460),D^{\prime}_{1}(2430))\mun\neumb$  (collectively referred to as $\Bb \to D^{**}(\to \Dstarp\pi)\mun\neumb$), form factors are
taken from Ref.~\cite{Leibovich:1998vw}.
The slope of the Isgur-Wise function~\cite{Isgur1989113,Isgur1990527} is included as a free parameter in the fit, with a constraint derived from fitting the $\Dstarp\mun\pim$ control sample. This fit also serves to validate this choice of model for this background. 
Contributions 
from $\Bsb \to (D^{\prime+}_{s1}(2536), D^{*+}_{s2}(2573))\mun\neumb$ decays use a similar parameterization, keeping only the lowest-order terms. 
Semileptonic decays to heavier charmed hadrons decaying as $D^{**} \to \Dstarp\pi\pi$ and semitauonic decays $\Bb \to (D^{}_{1}(2420),D^{*}_{2}(2460),D^{\prime}_{1}(2430))\taum\neut$
are modeled using the ISGW2\cite{ISGW2} parameterization. To improve the modeling for the former, a fit is performed to the $\Dstarp \mun \pip \pim$ control sample to generate an
 empirical correction to the \qq distribution, as the resonances that contribute to this final state and their respective form factors are not known. The contribution of semimuonic decays to excited charm states amounts to approximately 12\% of the
 normalization mode in the fit to the signal sample. 
 
 An important background source is \B decays into final states containing two charmed hadrons, $\Bb\to \Dstarp H_c X$, followed by 
 semileptonic decay of the  charmed hadron $H_c\to \mu\neum X$. This process occurs at 
 a total rate of 6\%--8\% relative to the normalization mode. The template for this process is generated using a
 simulated event sample of $B^+$ and $B^0$ decays, with an appropriate admixture of final states. Corrections to the simulated template
 are obtained by fitting the $\Dstarp\mun K^{\pm}$ control sample. A similar simulated sample is also used to generate kinematic distributions for
final states containing a tertiary muon from $\Bb\to \Dstarp\D^{-}_s X$ decays, with $D^{-}_s\to \taum\neutb$ and $\taum\to\mun\neumb\neut$. 

The kinematic distributions of hadrons misidentified as muons
are derived based on the sample of $\Dstarp h^{\pm}$ candidates. 
Control samples of $\Dstarp$ ($\Lz$) decays are used to determine the probabilities for a $\pi$ or $K$ ($p$) to be misidentified as a muon, and to generate a $3\times 3$ matrix of probabilities for each species to satisfy the criteria for identification as a $\pi,K$ or $p$. These are used to determine the composition of the $\Dstarp h^{\pm}$ sample in order to model the background from hadrons misidentified as muons. Two methods are developed to handle the unfolding of the individual contributions of $\pi$, $K$, and $p$, which result in different values for $\RDst$. The average of the two methods is taken as the nominal central value, and half the difference is assigned as a systematic uncertainty.

Combinatorial backgrounds are classified based on whether or not a genuine $\Dstarp \to \Dz\pip$ decay is present. 
Wrong-sign $\Dz\pim\mun$ 
combinations are used 
to determine the component with misreconstructed or false \Dstarp candidates. The size of this contribution is constrained by fitting the $\Delta m$ distribution of $\Dstarp\mun$ candidates in the full $\Delta m$ region.
The contribution from correctly reconstructed \Dstarp candidates combined with $\mun$ from unrelated $b$ hadron decays is determined from wrong-sign $\Dstarp\mup$ 
combinations. The size of this contribution is constrained by use of the mass region $5280 < m(\Dstarp\mu^{\mp}) < 10000 \mevcc$, which determines the expected ratio of $\Dstarp\mun$ to $\Dstarp\mup$ yields. In both cases, the contributions of misidentified muons are subtracted when generating the kinematic distributions for the fit.

The binned \mmsq, \El, and \qq distributions in data are fit using a maximum likelihood method with three dimensional templates representing the signal, the normalization and the background sources. To avoid bias, the procedure is developed and finalized without knowledge of the resulting value of \RDst. The templates extend over the kinematic region $-2 < \mmsq < 10  \gevgevcccc$ in 40 bins, $100 < \El < 2500 \mev$ in 30 bins, and $-0.4 < \qq < 12.6 \gevgevcccc$ in 4 bins. The fit extracts: the relative contributions of signal and normalization modes and their form factors; the relative yields of each of the $\Bb\to D^{**}(\to\Dstarp\pi)\mu\nu$ and their form factors; the relative yields of $\Bsb\to D_{s}^{**+}(\to\Dstarp\KS)\mun\neumb$ and $\Bb \to D^{**}(\to\Dstarp\pi\pi)\mun\neub$ decays; the relative yield of $\Bb\to \Dstarp H_{c}(\to\mu\nu X^{\prime})X$ decays; the yield of misreconstructed $\Dstarp$ and combinatorial backgrounds; and the background yield from hadrons misidentified as muons separately 
above and below $|p_\mu| = 10\gev$.
Uncertainties in the shapes of the templates due to the finite number of simulated events, which are therefore uncorrelated bin-to-bin, are incorporated directly 
into the likelihood using the Beeston-Barlow `lite' procedure \cite{Barlow:1993hy}. The fit includes shape uncertainties with bin-to-bin correlations ({\it e.g.} form factor uncertainties) 
via interpolation between nominal and alternative histograms. Control samples for partially reconstructed backgrounds (\ie $\Dstarp\mun\pim$, $\Dstar\mun\pip\pim$, and $\Dstar\mun K^{\pm}$)
are fit independently from the fit to the signal sample. Since the selections used for these control samples include inverting the isolation requirement used to select the signal sample, this method allows for the determination of the corrections to the  
$\Bb\to \Dstarp H_c (\to \mu\nu X^{\prime}) X$ and 
$\Bb \to \Dstarp \pi \pi \mun \neumb$ backgrounds with negligible influence from the signal and normalization events. The results are validated with an independently-developed alternative fit.  
 In this second approach, control samples are fit simultaneously with the signal sample with correction parameters allowed to vary, allowing correlations among parameters to be 
 incorporated exactly. This fit also forgoes the use of interpolation in favor of reweighting the simulated samples and recomputing the kinematic distributions for each value of the corresponding parameters. The two fits are extensively cross-checked and give consistent results.

\begin{figure}[p]
\begin{center}
\begin{overpic}[ scale=0.8]{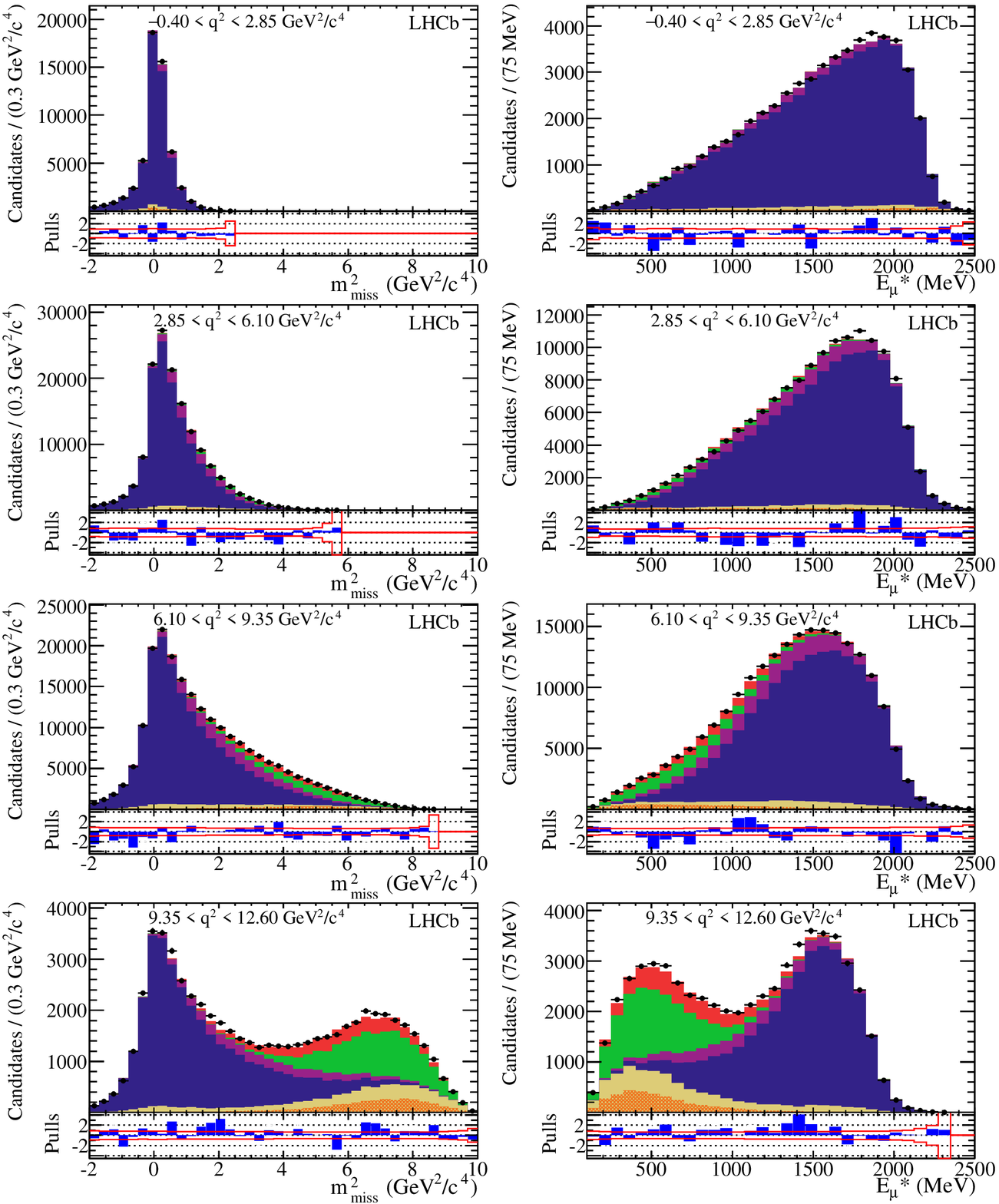}
	\put(18,83.7){\includegraphics[scale=0.175]{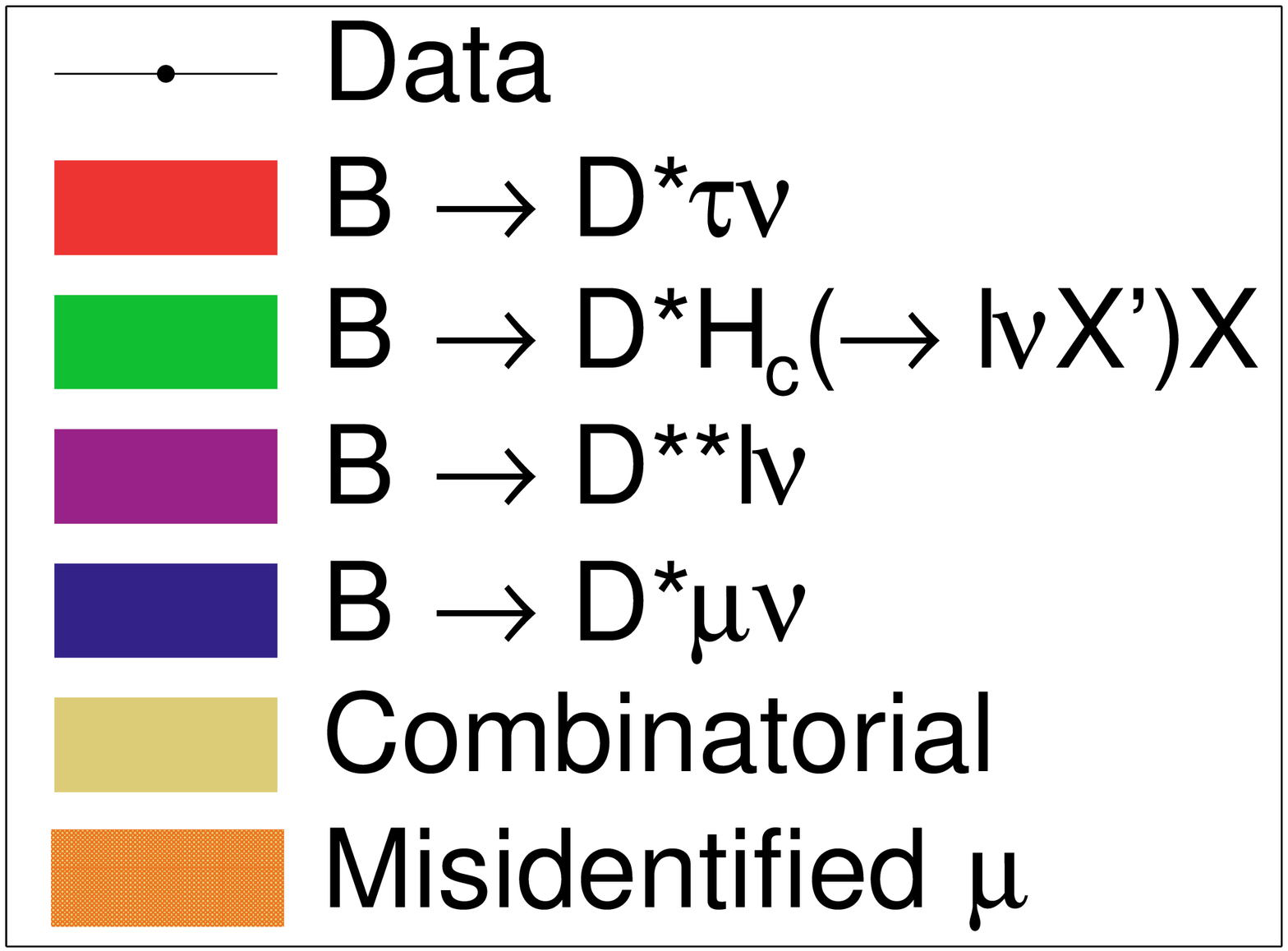}}
	\end{overpic}
\end{center}
\caption{Distributions of $m^2_{\rm miss}$ (left) and \El (right) of the four \qq bins of the signal data, overlaid with projections of the fit model with all normalization and shape parameters at their best-fit values. Below each panel differences between the data and fit are shown, normalized by the Poisson uncertainty in the data. The bands give the $1\sigma$ template uncertainties.}
\label{fig:theFit}
\end{figure}

The results of the fit to the signal sample are shown in Fig. \ref{fig:theFit}. 
Values of the \normmode form factor parameters determined by the fit agree with the current world average values. The fit finds $363~000\pm1600$ \normmode decays in the signal sample and an uncorrected ratio of yields $N(\sigmode)/N(\normmode) = (4.54\pm 0.46) \!\times\! 10^{-2}$. Accounting for the $\taum \to \mun\neumb\neut$ branching fraction \cite{PDG2014}
and the ratio of efficiencies results in $\mathcal{R}(\Dstar)=0.336\pm 0.034$, where the uncertainty includes the statistical uncertainty, the uncertainty due to form factors, and the statistical uncertainty in the kinematic distributions used in the fit. As the signal yield is large, this uncertainty is dominated by the determination of various background yields in the fit and their correlations with the signal, which are as large as $-0.68$ in the case of $\Bb\to \Dstarp H_{c}(\to\mu\nu X^{\prime})X$.

\begin{table}[b]
\begin{center}
\caption{Systematic uncertainties in the extraction of $\mathcal{R}(\Dstar)$.}
\label{tab:systematics}
\begin{tabular}{l r}\hline\hline
{\bf Model uncertainties} & {\bf Absolute size ($\mathbf{\times 10^{-2}}$})\\\hline
Simulated sample size & 2.0\\
Misidentified $\mu$ template shape & 1.6\\
$\Bzb\to\Dstarp(\taum/\mun)\neub$ form factors & 0.6\\
$\Bb \to \Dstarp H_{c}(\to \mu\nu X^{\prime})X$ shape corrections & 0.5\\
$\mathcal{B}(\Bb\to D^{**}\taum\neutb)/\mathcal{B}(\Bb\to D^{**}\mun\neumb)$ & 0.5\\
$\Bb\to D^{**}(\to \Dstar\pi\pi)\mu\nu$ shape corrections& 0.4\\
Corrections to simulation & 0.4\\
Combinatorial background shape & 0.3\\
$\Bb \to \D^{**}(\to\Dstarp\pi)\mun\neumb$ form factors & 0.3\\
$\Bb\to \Dstarp(\D_s\to \tau\nu)X$ fraction & 0.1\\
{\bf Total model uncertainty} & {\bf 2.8}\\\hline
{\bf Normalization uncertainties} & {\bf Absolute size ($\mathbf{\times 10^{-2}}$)} \\\hline
Simulated sample size & 0.6\\
Hardware trigger efficiency & 0.6\\
Particle identification efficiencies & 0.3\\
Form-factors & 0.2\\ 
$\mathcal{B}(\taum \to \mun\neumb\neut)$ & $<0.1$\\
{\bf Total normalization uncertainty} & {\bf 0.9}\\\hline
{\bf Total systematic uncertainty } & {\bf 3.0}\\\hline\hline
\end{tabular}
\end{center}
\end{table}

Systematic uncertainties on \RDst are summarized in Table~\ref{tab:systematics}. The uncertainty in extracting \RDst from the fit (model uncertainty) is dominated by the 
statistical uncertainty of the simulated samples; this contribution is estimated via the reduction in the fit uncertainty when the sample statistical uncertainty is not considered in the likelihood. The systematic uncertainty from the kinematic shapes of the background from hadrons misidentified as muons is taken to be half the difference in \RDst using the two unfolding methods. 
Form factor parameters are included in the likelihood as nuisance parameters, and represent a source of systematic uncertainty. The total uncertainty on \RDst estimated from the fit therefore incorporates these sources. To separate the statistical uncertainty and the contribution of the form factor uncertainty, the fit is repeated with form factor parameters fixed to their best-fit values, and the reduction in uncertainty is used to determine the contribution from the form factor uncertainties.
The systematic uncertainty from empirical corrections to the kinematic distributions of $\Bb \to D^{**}(\to \Dstarp\pi\pi)\mun\neumb$ and $\Bb \to \Dstarp H_{c}(\to\mu\nu X^{\prime})X$ backgrounds is also computed based on fixing the relevant parameters to their best fit values, as described above. 
The contribution of 
$\Bb \to D^{**}(\to\Dstarp\pi)\taum\neutb$, $\Bb \to D^{**}(\to\Dstarp\pi\pi)\taum\neutb$ and $\Bsb \to (D^{+}_{s1}(2536), D^{+}_{s2}(2573))\taum\neutb$ 
events is fixed to $12\%$ of the corresponding semimuonic modes, with half of this yield assigned as a 
systematic uncertainty on \RDst. Similarly the contribution of $\Bb \to \Dstarp D^{-}_s (\to \taum \neutb)$ decays is fixed 
using known branching fractions \cite{PDG2014}, and $30\%$ changes in the nominal value are taken as a systematic uncertainty. 
Corrections to the modeling of variables related to the pointing of the \Dz candidates to the PV are needed to derive the kinematic distributions for the fit. These corrections are derived from a comparison of simulated \normmode events with a pure \normmode data sample, and a systematic 
uncertainty is assigned by computing an alternative set of corrections using a different selection for this data 
subsample.

The expected yield of $\Dstarp\mun$ candidates compared to $\Dstarp\mup$ candidates (used to model the 
combinatorial background) varies as a function of $m(\Dstarp\mu^{\mp})$. The size of this effect is estimated in the $5280 < m(\Dstarp\mu^{\mp}) < 10000 \mevcc$ region and the uncertainty is propagated as a systematic uncertainty on \RDst.

Uncertainties in converting the fitted ratio of signal and normalization yields into \RDst (normalization uncertainties) come from the finite statistical precision of the simulated samples used to determine the efficiency ratio, and several other sources. The efficiency of the hardware triggers obtained in simulation differs between magnet polarities and between \pythia versions --- the midpoint of the predictions is taken as the nominal value and the range of variation is taken as a systematic uncertainty on the efficiency ratio. Particle identification efficiencies are applied to simulation based on binned $\jpsi \to \mup \mun$ and $\Dz \to \Km\pip$ control samples, which introduces a systematic uncertainty that is estimated by binning the control samples differently and by comparing to simulated particle identification. The signal and normalization form factors alter the expected ratio of detector acceptances, and $1 \sigma$ variations in these with respect to the world averages are used to to assign a 
systematic 
uncertainty. Finally, the uncertainty in the current world average value of $\mathcal{B}(\taum\to\mun\neumb\neut)$ contributes a small normalization uncertainty.

In conclusion, the ratio of branching fractions $\RDst=\mathcal{B}(\Bzb\to D^{*+}\taum\neutb)/\mathcal{B}(\Bzb\to D^{*+}\mun\neumb)$ is measured to be $0.336 \pm 0.027\stat \pm 0.030\syst$. 
The measured value is in good agreement with previous measurements at \babar and \belle \cite{Matyja:2007kt, Bozek:2010xy} and is $2.1$ standard deviations greater than the SM expectation of $0.252 \pm 0.003$~\cite{Fajfer:2012dp}.
This is the first measurement of any decay of a \bquark hadron into a final state with tau leptons at a hadron collider, and the techniques demonstrated in this letter open the possibility to study a broad range of similar $b$ hadron decay modes with multiple missing particles in hadron collisions in the future.

We express our gratitude to our colleagues in the CERN
accelerator departments for the excellent performance of the LHC. We
thank the technical and administrative staff at the LHCb
institutes. We acknowledge support from CERN and from the national
agencies: CAPES, CNPq, FAPERJ and FINEP (Brazil); NSFC (China);
CNRS/IN2P3 (France); BMBF, DFG, HGF and MPG (Germany); INFN (Italy); 
FOM and NWO (The Netherlands); MNiSW and NCN (Poland); MEN/IFA (Romania); 
MinES and FANO (Russia); MinECo (Spain); SNSF and SER (Switzerland); 
NASU (Ukraine); STFC (United Kingdom); NSF (USA).
The Tier1 computing centres are supported by IN2P3 (France), KIT and BMBF 
(Germany), INFN (Italy), NWO and SURF (The Netherlands), PIC (Spain), GridPP 
(United Kingdom).
We are indebted to the communities behind the multiple open 
source software packages on which we depend. We are also thankful for the 
computing resources and the access to software R\&D tools provided by Yandex LLC (Russia).
Individual groups or members have received support from 
EPLANET, Marie Sk\l{}odowska-Curie Actions and ERC (European Union), 
Conseil g\'{e}n\'{e}ral de Haute-Savoie, Labex ENIGMASS and OCEVU, 
R\'{e}gion Auvergne (France), RFBR (Russia), XuntaGal and GENCAT (Spain), Royal Society and Royal
Commission for the Exhibition of 1851 (United Kingdom).

\clearpage

\addcontentsline{toc}{section}{References}
\setboolean{inbibliography}{true}
\bibliographystyle{LHCb}
\bibliography{main,LHCb-PAPER,LHCb-CONF,LHCb-DP,LHCb-TDR,SL}

\newpage

\newpage
%%%%%%%%%%%%%%%%%%%%%%%%%%%%%%%%%%%%%%%%%%
\centerline{\large\bf LHCb collaboration}
\begin{flushleft}
\small
R.~Aaij$^{38}$, 
B.~Adeva$^{37}$, 
M.~Adinolfi$^{46}$, 
A.~Affolder$^{52}$, 
Z.~Ajaltouni$^{5}$, 
S.~Akar$^{6}$, 
J.~Albrecht$^{9}$, 
F.~Alessio$^{38}$, 
M.~Alexander$^{51}$, 
S.~Ali$^{41}$
% ...

{\footnotesize \it
% ...
$ ^{5}$Clermont Universit\'{e}, Universit\'{e} Blaise Pascal, CNRS/IN2P3, LPC, Clermont-Ferrand, France\\
$ ^{6}$CPPM, Aix-Marseille Universit\'{e}, CNRS/IN2P3, Marseille, France\\
% ...
$ ^{9}$Fakult\"{a}t Physik, Technische Universit\"{a}t Dortmund, Dortmund, Germany\\
% ...
$ ^{37}$Universidad de Santiago de Compostela, Santiago de Compostela, Spain\\
$ ^{38}$European Organization for Nuclear Research (CERN), Geneva, Switzerland\\
% ...
$ ^{41}$Nikhef National Institute for Subatomic Physics, Amsterdam, The Netherlands\\
% ...
$ ^{46}$H.H. Wills Physics Laboratory, University of Bristol, Bristol, United Kingdom\\
% ...
$ ^{51}$School of Physics and Astronomy, University of Glasgow, Glasgow, United Kingdom\\
$ ^{52}$Oliver Lodge Laboratory, University of Liverpool, Liverpool, United Kingdom\\
% ...
}
\end{flushleft}
%%%%%%%%%%%%%%%%%%%%%%%%%%%%%%%%%%%%%%%%%%

\end{document}